\documentclass{aa}  
\usepackage[colorlinks=true]{hyperref}
\hypersetup{citecolor= blue, linkcolor=red, urlcolor=blue}
\usepackage[normalem]{ulem}

\newcommand{\dd}{\mathrm{d}}

\usepackage{multirow}

\usepackage{graphicx}
\usepackage{txfonts}
\begin{document}

   \title{Radiative losses in the chromosphere during a C-class flare}

   \author{Rahul Yadav
          \inst{1,2}
          \and
          J. de la Cruz Rodr\'iguez
          \inst{1}
          \and
          Graham S. Kerr
          \inst{3,4}
          \and 
          C. J. D\'iaz Baso
          \inst{1}
          \and 
          Jorrit Leenaarts
          \inst{1}
          }

   \institute{Institute for Solar Physics, Dept. of Astronomy, Stockholm University, AlbaNova University Centre, SE-10691 Stockholm, Sweden 
   \and Laboratory for Atmospheric and Space Physics, University of Colorado, 3665 Discovery Drive, Boulder, CO 80303, USA\\ \email{rahul.yadav@lasp.colorado.edu}
   \and Department of Physics, Catholic University of America, 620 Michigan Avenue, Northeast, Washington, DC 20064, USA
   \and NASA Goddard Space Flight Center, Heliophysics Sciences Division, 8800 Greenbelt Road, Greenbelt, MD 20771, USA
}
   \date{Draft: compiled on \today%\ at \currenttime~UT
   }

% \abstract{}{}{}{}{} 
% 5 {} token are mandatory
 
  \abstract
   {Solar flares release an enormous amount of energy ($\sim$10$^{32}$~erg) into the corona. A substantial fraction of this energy is transported to the lower atmosphere, which results in chromospheric heating. The mechanisms that transport energy to the lower solar atmosphere during a flare are still not fully understood. }
   {We aim to estimate the temporal evolution of the radiative losses in the chromosphere at the footpoints of a C-class flare, in order to set observational constraints on the electron beam parameters of a RADYN flare simulation.}
   {We estimated the radiative losses from hydrogen, and singly ionized Ca and Mg using semiempirical model atmospheres, which were inferred from a multiline inversion of observed Stokes profiles obtained with the CRISP and CHROMIS instruments on the Swedish 1-m Solar Telescope. The radiative losses were computed taking into account the effect of partial redistribution and non-local thermodynamic equilibrium. 
   To estimate the integrated radiative losses in the chromosphere, the net cooling rates were integrated between the temperature minimum and the height where the temperature reaches 10~kK. We also compared our time series of radiative losses with those from the RADYN flare simulations.}
   {We obtained a high spatial-resolution map of integrated radiative losses around the flare peak time. The stratification of the net cooling rate suggests that the Ca~IR triplet lines are responsible for most of the radiative losses in the flaring atmosphere. 
   During the flare peak time, the contribution from \ion{Ca}{ii}~H~\&~K and \ion{Mg}{ii}~h \&~k lines are strong and comparable to the Ca IR triplet ($\sim$32~kW~m$^{-2}$). Since our flare is a relatively weak event, the chromosphere is not heated above 11~kK, which in turn yields a subdued Ly$\alpha$ contribution ($\sim$7~kW~m$^{-2}$) in the selected limits of the chromosphere. 
  The temporal evolution of total integrated radiative losses exhibits sharply rising losses (0.4~kW~m$^{-2}$~s$^{-1}$) and a relatively slow decay (0.23~kW~m$^{-2}$~s$^{-1}$). The maximum value of total radiative losses is reached around the flare peak time, and can go up to 175~kW~m$^{-2}$ for a single pixel located at footpoint. After a small parameter study, we find the best model-data  consistency in terms of the amplitude of radiative losses and the overall atmospheric structure with a RADYN flare simulation in the injected energy flux of $5\times10^{10}$~erg~s$^{-1}$~cm$^{-2}$.}
{}
\maketitle

%
%-------------------------------------------------------------------

\section{Introduction}
Solar flares are the most energetic events on the Sun. A significant amount of energy, up to 10$^{32}$~erg, is released during an intense solar flare. 
A typical flare is characterized by a rapid increase in emission over a wide range of the electromagnetic spectrum, which affects the entire solar atmosphere, namely the photosphere, the chromosphere, and the corona \citep{2011SSRv..159...19F}. In the standard flare model \citep{1964NASSP..50..451C,1966Natur.211..695S,1974SoPh...34..323H,1976SoPh...50...85K}, the fundamental mechanism of solar flares is believed to be magnetic reconnection in the corona, converting the magnetic energy into kinetic and thermal energy. Following energy release in the corona, a substantial amount of energy is also transported to the lower solar atmosphere, resulting in intense plasma heating and ionization, ultimately leading to a substantial increase in the local radiative output in flare ribbons and footpoints. Observations have demonstrated that during flares, a major amount of the energy budget is radiated in the chromosphere \citep{2017LRSP...14....2B}, making understanding the flaring chromosphere key to understanding flare energetics.

There is unambiguous evidence of the presence of energetic particles in solar flare footpoints, and a large body of work points to the prominent role of directed beams of nonthermal electrons as a significant mechanism by which energy is transported from the corona to the chromosphere \citep[][]{1971SoPh...18..489B,1972SoPh...24..414H,2011SSRv..159..301K,2011SSRv..159..107H}. However, it is not known to what extent other energy transport mechanisms play a role. Examples include direct in situ heating at loop tops, which is conducted through to the lower atmosphere, the acceleration of protons and heavier ions alongside electrons, or the presence of flare-induced magnetohydrodynamic waves \citep[e.g.,][]{1982SoPh...80...99E,2008ApJ...675.1645F,2016ApJ...827..101K,2018ApJ...853..101R}. Determining the relative contributions of each of these mechanisms is an important goal in flare physics, which requires, in part, a solid understanding of the observed radiative losses in the lower atmosphere.

Semiempirical models have been used to estimate the radiative losses in the lower solar atmosphere by various authors \citep{1975SoPh...42..395M,1980ApJ...242..336M,1981ApJS...45..635V,1986lasf.conf..216A,1993ApJ...416..886G,2016ApJ...826...49S, 2020ApJ...890...22A}. 
These models have been performed using some form of non-local thermodynamical equilibrium (NLTE) data inversion. Inversion codes adjust the stratification of physical parameters in a model atmosphere in order to fit the observed spectra. The resulting model can be assumed to be an approximation of the real thermodynamical state of the solar atmosphere when the code includes all the physical ingredients that are required to model the observations \citep{2016LRSP...13....4D}.

The radiative losses estimated from commonly used semiempirical models of the quiet-Sun chromosphere are 4.3~kW~m$^{-2}$, whereas in active regions, the value is 20~kW~m$^{-2}$ \citep{1977ARA&A..15..363W, 1981ApJS...45..635V}. 
In order to maintain the energy balance in the chromosphere, some form of heating mechanism is required, which is still under debate. 
 Comparison of observations with the semiempirical models has revealed that the chromospheric radiative losses are mainly dominated by strong lines of \ion{Ca}{ii}, \ion{Mg}{ii}, Ly$\alpha$, and the H$^{-}$ continuum \citep{1981ApJS...45..635V}. 
 \cite{1989ApJ...346.1010A} studied the role of \ion{Fe}{ii} lines in the chromosphere and concluded that \ion{Fe}{ii} can be responsible for up to one half of the radiative losses if they are included in the calculations from quiet-Sun models. Despite this potentially important contribution, \ion{Fe}{ii} lines have not been traditionally included in the calculation of the radiative losses because of the vast number of transitions that would need to be calculated in NLTE, and the lack of accurate atomic models.
 Semiempirical models have also been used to investigate various possible energy-transfer processes occurring during flares \citep{1980ApJ...242..336M}.

\cite{2018A&A...612A..28L} used the \ion{Ca}{ii}~K brightness as a
proxy of the radiative losses in the chromosphere in order to investigate the atmospheric structure and heating of the solar chromosphere in an emerging flux region. They reported that the frequency-integrated \ion{Ca}{ii}~K brightness correlates strongly with the total linear polarization in the \ion{Ca}{ii}~8542 \AA\ line, while the Ca II K profile shapes indicate that the bulk of the radiative losses occurs in the lower chromosphere. In a recent study, \cite{2021A&A...647A.188D} estimated radiative losses from the inferred atmospheric model in a strong magnetic reconnection event by integrating the net cooling rates in the chromosphere from the temperature minimum to the height where the temperature reaches
10~kK. They reported integrated radiative losses up to 160 kW~m$^{-2}$, which is roughly five times more than the
typical value for the chromosphere in active regions.

Due to recent improvements in solar observations and in inversion techniques, it is now possible to infer the stratification of physical parameters such as temperature, line-of-sight velocity, micro-turbulence velocity, and magnetic field in the solar atmosphere using multispectral line observations \citep{2019A&A...623A..74D, 2019A&A...627A.101V}. Recently, \citet[hereafter Paper~I]{2021A&A...649A.106Y} constructed the stratification of physical parameters in a flaring atmosphere using multiple spectral lines forming in different parts of the solar atmosphere.
Such multiline observations, spanning different layers of the chromosphere and photosphere, allow us to reconstruct the atmospheric parameters more accurately compared to a single layer or spectral line. 

The temporal analysis of the inferred parameters in Paper~I revealed that the mean temperature increased up to $\sim$11~kK in the chromosphere (between $\log\tau_{500}\sim-3.5$ and $-5$), whereas no significant change was noticed in the photosphere. As a consequence, intense brightening, which is a signature of radiative losses, is observed in the chromospheric lines (\ion{Ca}{ii}~K and \ion{Ca}{ii}~8542~\AA) at flare ribbons.

In this paper, we estimated the radiative losses from  hydrogen, singly ionized Ca, and singly ionized Mg using the semiempirical flare model inferred from multiline observations. We report the temporal evolution of the integrated radiative losses in the chromosphere during the flare. 
We also compared our results with radiative losses calculated from a flare simulation performed with the RADYN code \citep{1997ApJ...481..500C, 2015ApJ...809..104A} in order to demonstrate how our observationally derived results can attempt to set  constraints on the energy injection. 

A brief overview of observations and inversions is given in Section \ref{Sec_observation}. The approach for estimating the cooling rates is described in Section \ref{sec_rates_intro}. The results are discussed in Section \ref{Sec_results}. Finally, the paper is summarized in Section \ref{Sec_summary}.

\begin{figure*}[!h]
    \centering
    
    \includegraphics[width=0.58\linewidth]{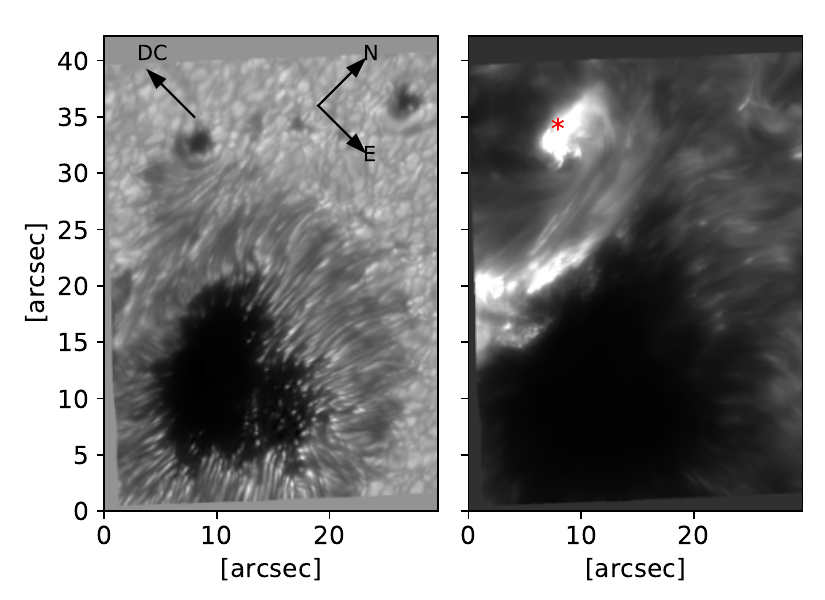} \hspace*{-0.1in}
    \includegraphics[width=0.405\linewidth]{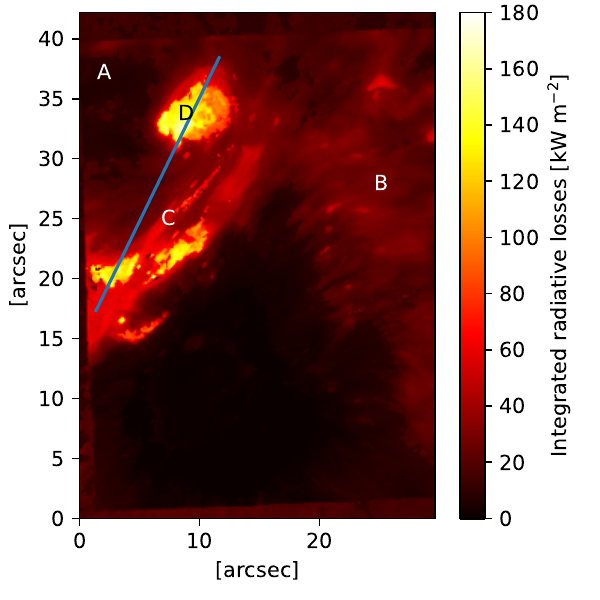}
    \caption{Overview of the flare observed at 08:46~UT on May 6, 2019. The continuum intensity map (\textit{left panel}) and the Chromospheric intensity map (\textit{middle panel}) near the line core of \ion{Ca}{ii}~K, observed using the SST. \textit{Right panel}: Integrated radiative losses estimated from the inferred semiempirical model atmosphere.
    Solar north, solar east, and the direction of the disk center are indicated by ``N,'' ``E,'' and ``DC,'' respectively. The red asterisk symbol in the middle panel shows the location of pixels analyzed to investigate the temporal evolution of the integrated radiative losses. The mean values of the total integrated losses at the A, B, C, and D locations (5$\times$5 pixels) are listed in Table \ref{Table1}.} 
    \label{fig:overview_fig}
\end{figure*}

%--------------------------------------------------------------------
\section{Observation and inversion}
\label{Sec_observation}
A C2-class solar flare (SOL2019-05-06T08:47) was observed between 08:34 and 9:33 UT on May 6, 2019, with the CRisp Imaging SpectroPolarimeter \citep[CRISP;] []{Scharmer2008} and the CHROMospheric Imaging Spectrometer \citep[CHROMIS;][]{2017psio.confE..85S} instruments on the Swedish 1-m Solar Telescope \citep[SST;][]{2003SPIE.4853..341S}. This flare occurred in the active region (AR) NOAA 12740, which was located at N08E48 ($\mu$=0.62). The CRISP simultaneously recorded full spectropolarimetric data in the \ion{Ca}{ii}~8542~\AA\ and \ion{Fe}{i}~6173 \AA\ spectral lines. The \ion{Ca}{ii} 8542 \AA\ line scans consisted of 17 wavelength positions spanning a range of 1.4~\AA\ around line center, with steps of 75~m\AA\ in the inner wings and 125~m\AA~at four outer wing positions, whereas the \ion{Fe}{i}~6173~\AA\ spectral line was scanned at 15 wavelength positions spanning a range of 0.5~\AA\ around line center, with steps of 35 m\AA~in the inner wings and 50 m\AA\ in the outer wings at four different positions. The CHROMIS recorded \ion{Ca}{ii}~K intensity profiles at 28 wavelength positions spanning a range of 3~\AA\ around line center, with steps of 65 m\AA\ in the inner wings and an additional sampling at wavelength positions $\pm{1.5}$~\AA\ and $\pm{1.18}$~\AA\ relative to the line center. In addition to this, one point in the continuum at 4000~\AA\ was also observed with the CHROMIS instrument. The CRISP and CHROMIS data were obtained quasi-simultaneously with a cadence of 21~sec and 15~sec, respectively. 

The data were reduced and post-processed using the CRISPRED/SSTRED pipelines \citep{2015A&A...573A..40D,2021A&A...653A..68L}. Further observational details, data reduction, and post-processing procedures are described in Paper I. See Figure 2 in Paper~I for an overview of the observed flare at the SST, along with aligned-ultraviolet (UV) and extreme-ultraviolet (EUV) images observed by the Atmosphere Imaging Assembly (AIA; \citealt{2012SoPh..275...17L}) on board the Solar Dynamic Observatory (SDO;
\citealt{2012SoPh..275..207S}).

We employed the STiC inversion code \citep{delaCruz2016,2019A&A...623A..74D} to infer the atmospheric stratification of temperature, velocity, and magnetic field from the \ion{Fe}{i} 6173~\AA,\, \ion{Ca}{ii}~K, and \ion{Ca}{ii}~8542~\AA\ spectral lines. The STiC code is built around a modified version of the RH code \citep{Uitenbroek2001} in order to derive the atomic populations by assuming statistical equilibrium and a plane-parallel geometry. The equation of state is borrowed from the Spectroscopy Made Easy (SME) computer code described in \cite{2017A&A...597A..16P}. The radiative transport equation is solved using cubic Bezier solvers \citep{2013ApJ...764...33D}.

For the inversion of the Stokes profiles, we considered the \ion{Ca}{II}~8542~\AA\ line in NLTE conditions, under the assumption of complete frequency redistribution, while the \ion{Ca}{ii}~K line was synthesized in NLTE conditions with partial redistribution (PRD) effects of scattered photons. We used the fast approximation proposed by \cite{2012A&A...543A.109L}. The \ion{Fe}{I} 6173~\AA\ line was treated under the assumption of LTE conditions. We inverted all the Stokes parameters in the \ion{Fe}{I}~6173~\AA\ and \ion{Ca}{II}~8542~\AA\ lines, but only Stokes $I$ in the \ion{Ca}{ii}~K line. The inversion setup and retrieved atmospheric parameters are discussed in detail in Paper~I.

%-----------------------------------------------------------------
\section{Radiative cooling rate}
\label{sec_rates_intro}
The energy released by radiation is
characterized by net radiative cooling rates (radiative losses), which are evaluated as the frequency-integrated radiative flux divergence \citep{2003rtsa.book.....R}:
\begin{equation}
    \nabla \cdot F = \int_{0}^{\infty} \int_{\Omega}^{} \alpha_{\nu} (S_{\nu} -I_{\nu}) \, \mathrm{d}\Omega \, \mathrm{d}\nu,
\label{loss_integral}
\end{equation}
where $\alpha$, $S$, and $I$ are the opacity, source function, and intensity at the frequency $\nu$ and in direction $\Omega$. 

Equation~\ref{loss_integral} describes the total radiative loss rate. In this study we focus on losses in individual spectral lines. The contribution of a single bound-bound transition to Eq.~~\ref{loss_integral}
can be expressed as
\begin{eqnarray}
Q_\mathrm{bb} &=& \oint \int \frac{h\nu}{4\pi} \Bigg[ n_j A_{ji} \psi_{ij}(\nu,\vec{n})   \nonumber \\
&& - B_{ij} \bigg( n_i - \frac{g_i}{g_j} \rho_{ij}(\nu,\vec{n}) n_j \bigg) \phi_{ij}(\nu,\vec{n}) I(\nu,\vec{n}) \Bigg]
\, \dd \nu \,  \dd \Omega, \label{eq:bb} 
\label{colling_equation}
\end{eqnarray}
where $h$ is the Planck constant, $n_j$ and $n_i$ are the population densities of the upper level and lower level $i$, $A_{ji}$ and $B_{ij}$ are Einstein coefficients, $g_i$ and $g_j$ are the statistical weights of level $i$ and $j$, and $I(\nu,\boldsymbol{n})$ is the intensity for a given direction $\vec{n}$ at frequency $\nu$.
This formulation takes PRD effects into account through the frequency and angle-dependent ratio of the emissivity to the absorption profile $\rho_{ij} = \psi_{ij}/\phi_{ij}$
\citep{2002ApJ...565.1312U}. 

We note that Eq.~\ref{loss_integral} includes many continuum opacity sources that are being neglected when we make a selection of transitions and apply Eq.~\ref{colling_equation}. In flares, many of those continua can cool or heat, even in the chromosphere. A limitation is that most of them are computed under the assumption of LTE in the RH module of STiC, and this assumption cannot be used to properly estimate atom population densities in the chromosphere. Therefore, we opted for a conservative approach, and we are only accounting for the contributions that we can model reasonably well. Therefore, our results pose a lower-limit estimate of the radiative losses.

The net cooling rates are evaluated from the inferred flare model by summing the losses in the Ly$\alpha$, \ion{Ca}{II}~H~\&~K, Ca IR triplet (\ion{Ca}{II}~8542, 8498, 8662~\AA), and \ion{Mg}{II} h~\&~k spectral lines as computed using Eq.~\ref{colling_equation}. The radiative rates and populations were computed using the STiC code after taking into account the PRD effect and NLTE conditions (see Section \ref{Sec_observation} for the STiC code and inversion). 

We decided to ignore the contribution from bound-free transitions in the integrated losses because the transition region is not well constrained in our inferred models, as the observed spectral lines form in the lower and middle chromosphere. Additionally, the node description of the atmosphere sets limits on the steepness of the transition region by imposing a relatively smooth stratification of the physical parameters with height \citep{delaCruz2016,2019A&A...623A..74D}. Because the code did not need to place a steep transition region in order to fit the \ion{Ca}{II} lines, the model predicts, close to the transition region, contributions from the Lyman and Balmer continua that are unrealistically high. Therefore, our calculated radiative losses are only a lower-limit to the real ones, which in turn are a lower-limit approximation to the heating terms in the chromosphere.  This is not a problem for our analysis as we  compute the radiative losses in the same way with the inversion models and the simulations.

Although our flare was not observed in the \ion{Mg}{II}~h~\&~k and \ion{Ca}{II}~H lines, we assumed that the inferred model atmosphere retrieved from \ion{Ca}{ii}~8542~\AA, \ion{Ca}{II}~K, and \ion{Fe}{I}~6173~\AA\ observations represent the upper chromosphere correctly near the formation height of \ion{Mg}{II}~h~\&~k \citep{2019A&A...631A..33B}. Our observed flare is a relatively weak event and does not influence the photosphere as white-light flares do. Therefore, we neglected the contribution from weak spectral lines in the photosphere and the H$^{-}$ continuum as their contributions in the lower and middle chromosphere are small compared to the \ion{Ca}{II} and \ion{Mg}{II} lines \citep{1981ApJS...45..635V}. 
We note that in our inferred model the mean temperature in the lower and middle chromospheres is normally below 11~kK, and the uncertainty in the parameters is larger in the higher layers (<~log${\tau_{500}}= -5$) where the observed lines are less sensitive. 

\begin{figure*}[!t]
\centering
\includegraphics[width=\columnwidth]{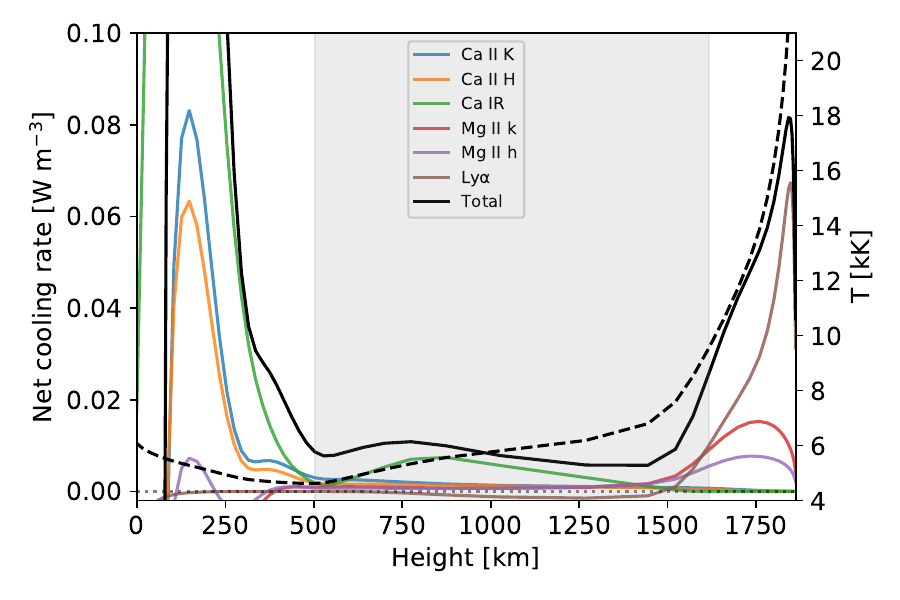}
\includegraphics[width=\columnwidth]{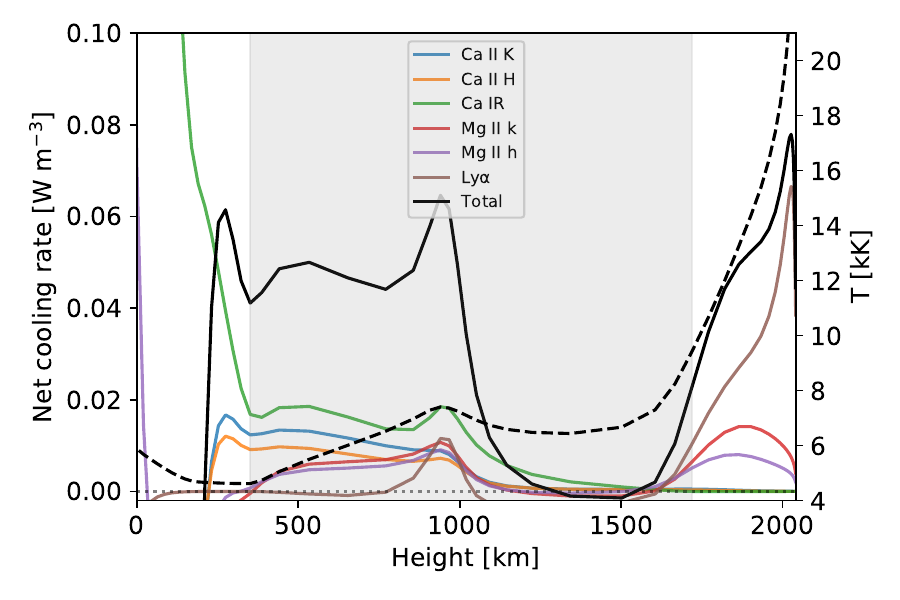}
\caption{\footnotesize{Stratification of the net radiative cooling rates per unit volume for a quiet-Sun pixel (left panel) and a pixel located at the flare footpoint (right panel) indicated by the asterisk symbol in Figure \ref{fig:overview_fig}. The solid colored lines refer to different spectral lines. The solid black  line represents the total cooling rate, whereas the dashed curve refers to the temperature stratification. The gray shaded area indicates the integrated heights.}}
\label{fig_loss_stratified}
\end{figure*}

\section{Results}
\label{Sec_results}
\subsection{Stratification and distribution of radiative losses}
We evaluated the bound-bound radiative losses per unit volume using the approach outlined in Section \ref{sec_rates_intro}. In order to estimate the total height-integrated radiative losses in the chromosphere, we followed the approach adopted by \citet{2021A&A...647A.188D}. We integrated the contribution of selected lines over geometrical heights, derived by assuming hydrostatic equilibrium using the STiC code, between the temperature minimum and the height where the temperature reaches 10 kK. As an example, the total integrated radiative losses for the field of view (FOV) comprising flare ribbons (near the flare peak time) are shown in Figure \ref{fig:overview_fig}. The figure illustrates that, during the flare peak time, the bright flare ribbon regions display strong integrated radiative losses.

The distribution of radiative losses across the FOV is greatly correlated with the integrated intensity of the \ion{Ca}{ii}~H\&K lines, similarly reported by \citet{2018A&A...612A..28L} and \citet{2021A&A...647A.188D}.
We note that the integrated radiative losses are strongly dependent on the heights used to integrate the net cooling rates. The contribution changes significantly in the deeper layers compared to higher layers as density drops sharply outward in the solar atmosphere. The stratification of the net radiative cooling rate per unit volume for a pixel located at the quiet-Sun and the flare footpoint are shown in Fig.~\ref{fig_loss_stratified}. The analysis of the quiet-Sun pixel exhibits that the contribution of \ion{Ca}{ii}~IR lines is dominating compared to the Ly$\alpha$, \ion{Ca}{II}~H~\&~K, and \ion{Mg}{ii}~h~\&~k lines. On the other hand, during the flare peak time, the net cooling rate at the flare footpoint shows that the contribution from the  \ion{Ca}{ii}~H~\&~K and \ion{Mg}{ii}~h \&~k lines is also strong and comparable to the Ca IR triplet ($\sim$32~kW~m$^{-2}$). The stratification of the net cooling rate of the Ly$\alpha$ line shows that it increases around 1~Mm but remains below that of the \ion{Ca}{ii}~IR lines.

To compare the integrated radiative losses at different parts of the FOV (near the flare peak time), we selected small patches of pixels (5$\times$5) indicated by A (the Quiet-Sun region), B (bright fibrils), C (near the polarity inversion line), and D (the flare footpoint) in the right panel of Fig.~\ref{fig:overview_fig}. The obtained integrated radiative losses and contribution of all selected lines at different locations are listed in Table \ref{Table1}. The total integrated value (4.9~kW~m$^{-2}$) in the Quiet-Sun region is compatible with previous findings \citep{1977ARA&A..15..363W, 1981ApJS...45..635V}. At the bright fibrils, the total  integrated value is 23.9~kW~m$^{-2}$. Near the polarity inversion line, the radiative losses are $\sim$37~kW~m$^{-2}$, which is roughly twice the value reported for active regions \citep{1981ApJS...45..635V}. At the flare footpoint, the mean integrated radiative loss value is $\sim$133~kW~m$^{-2}$. As listed in the table, in all cases, the majority contribution comes from the \ion{Ca}{II}~IR triplet lines, with the exception of the flare footpoint location, where the contribution from the \ion{Ca}{ii}~H~\&~K and \ion{Mg}{ii}~h~\&~k lines is similar to that from the \ion{Ca}{ii}~IR lines.

\begin{figure}[]
\centering
\includegraphics[width=0.42\textwidth]{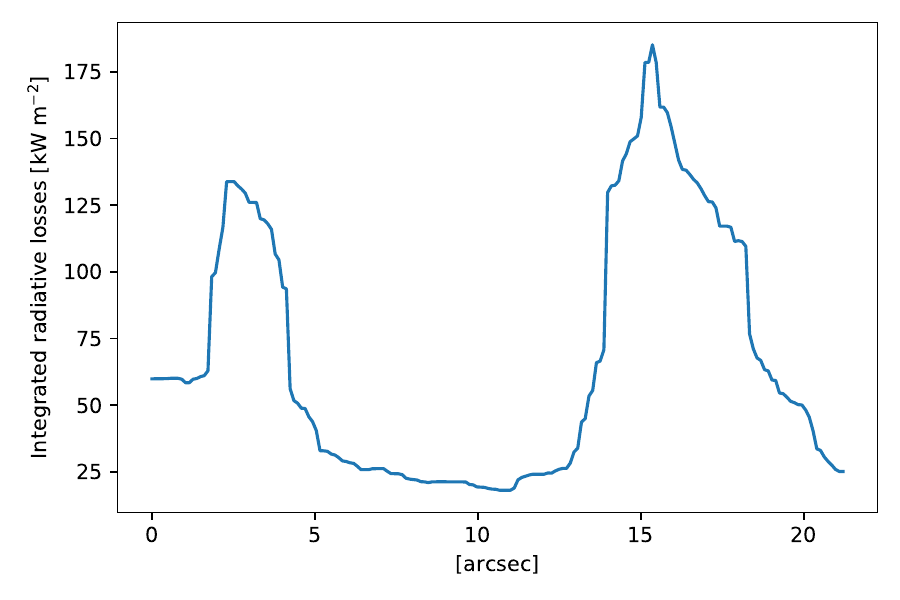}
\caption{\footnotesize{Total integrated radiative losses evaluated along a solid blue line passing through two footpoints shown in Fig.~\ref{fig:overview_fig}.}}
\label{integrated_line}
\end{figure}

Figure \ref{integrated_line} depicts the integrated radiative losses at the flare peak time, across a line passing through two footpoints. It shows that the maximum value can go up to $\sim$175~kW~m$^{-2}$ for pixels located at the flare footpoint. Similar high values are also reported by \cite{2021A&A...647A.188D} using a similar approach and spectral lines (except Ly$\alpha$) in a strong magnetic reconnection event. They report that the chromospheric radiative losses at the reconnection site are as high as 160~kW~m$^{-2}$.

\begin{figure}[]
\centering
\includegraphics[width=0.48\textwidth]{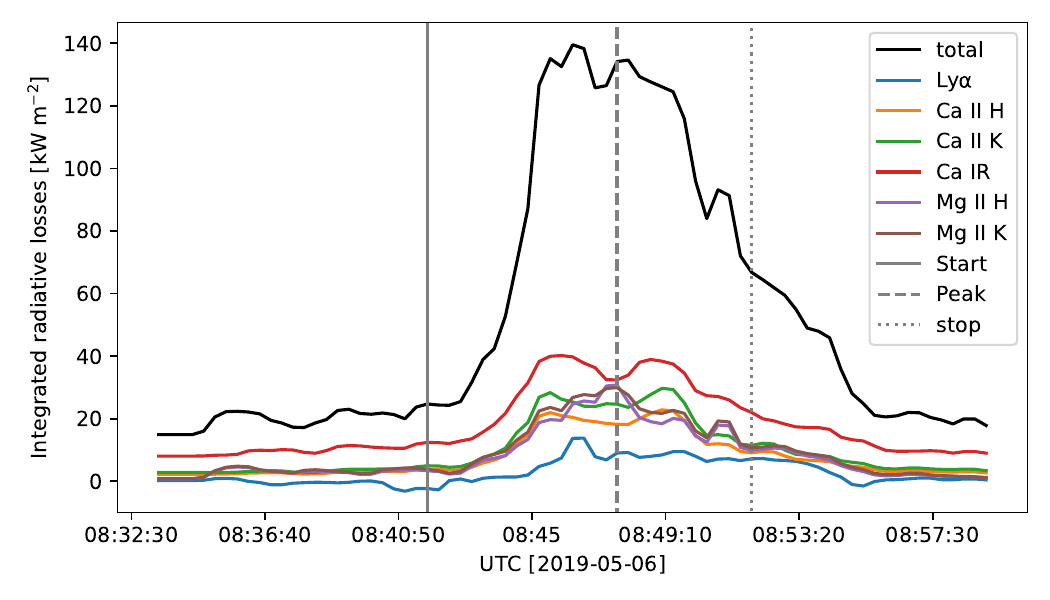}

\caption{Temporal evolution of the total integrated radiative losses in the chromosphere at the asterisk symbol highlighted in Fig.~\ref{fig:overview_fig}. The vertical solid, dashed, and dotted lines refer to the start, peak, and end times of the flare, respectively.} 
\label{integrated_footpoint}
\end{figure}

\begin{table}[!t]
\centering
\caption{Mean contribution of Ly$\alpha$, \ion{Ca}{ii}, and \ion{Mg}{ii} lines to the total integrated radiative losses (kW m$^{-2}$) for the regions highlighted in Fig.~\ref{fig:overview_fig}. The \ion{Ca}{}~IR refers to the triplet: \ion{Ca}{II}~8542, 8498, 8662~\AA. A, B, C, and D regions correspond to the quiet-Sun, bright fibrils, pixels near the polarity inversion line, and footpoint pixels at the flare peak time, respectively.}
\begin{tabular}{|c|cccc|}
\hline
\multirow{1}{*}{Spectral lines} & \multicolumn{4}{c|}{Regions} \\

       & A     & B    & C    & D     \\
\hline
\ion{Ca}{II}~H                         & 0.75  & 3.9  & 6.1  & 17.6  \\

\ion{Ca}{II}~K                         & 0.72  & 5.4  & 8.3  & 22.8  \\
Ca~IR                        & 3.2   & 9.1  & 14.9 & 32.5  \\
\ion{Mg}{II}~h                         & 0.27  & 2.4  & 3.4  & 25.2  \\
\ion{Mg}{II}~k                         & 0.21  & 2.8  & 3.7  & 28.2  \\
Ly$\alpha$  &  -0.17 & 0.3  & 0.76  &  7.05 \\
\hline
Total                           & 4.93   & 23.9 & 37.16 & 133.35 \\
\hline

\end{tabular}
\label{Table1}
\end{table}

\begin{figure*}[]
\centering
\includegraphics[width=0.55\textwidth]{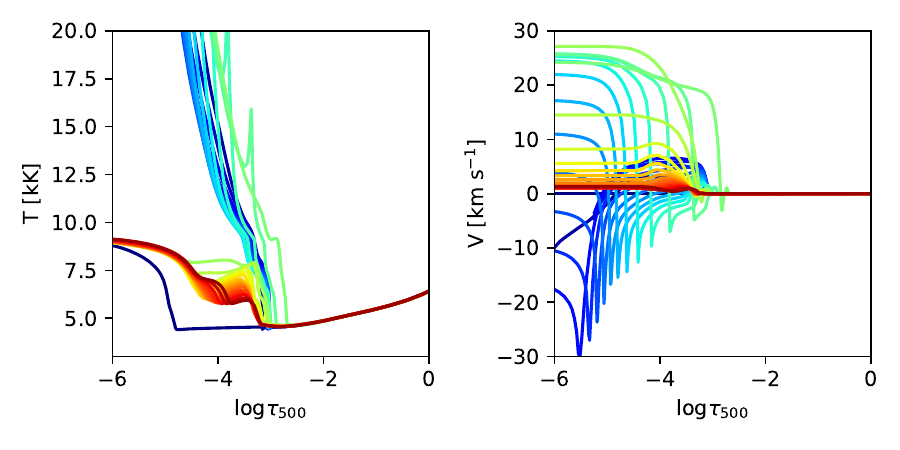}
\hspace{-4.5mm}
\raisebox{7mm}{\includegraphics[width=0.065\textwidth]{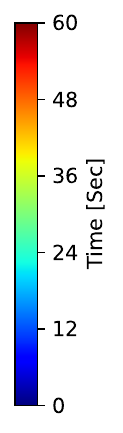}}
\includegraphics[width=0.35\textwidth]{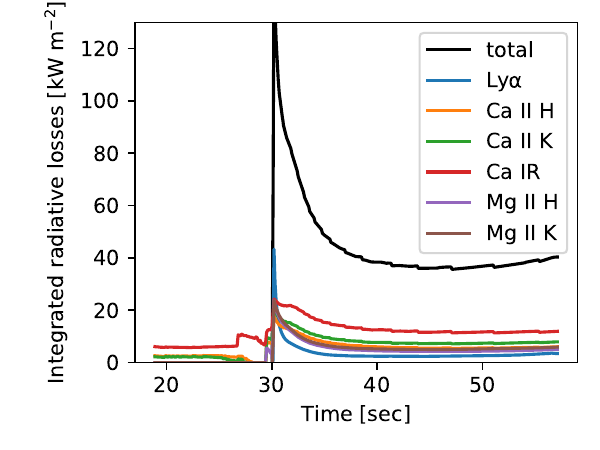}

\caption{Atmospheric stratification resulting from RADYN. \textit{Left and middle panels}: Temporal evolution of temperature and velocity in the RADYN simulation. The simulation was performed with $\delta=7$, E$_c=20$~keV and energy flux of $5\times10^{10}$ erg s$^{-1}$ cm$^{-2}$. \textit{Right panel}: Temporal evolution of the integrated radiative losses in the chromosphere estimated using the simulated atmosphere. Solid black lines refer to the total integrated radiative losses.}
\label{radyn_synthetic}
\end{figure*}

\subsection{Evolution of the total integrated radiative losses}
In order to investigate the temporal evolution of the integrated radiative losses at flare footpoints, we selected a small patch of pixels (5$\times$5) indicated by the asterisk symbol in the chromospheric intensity map of Fig.~\ref{fig:overview_fig}.
Figure \ref{integrated_footpoint} displays the obtained temporal evolution of the total integrated radiative losses at the flare footpoint. The vertical lines in Fig.~\ref{integrated_footpoint} indicate the flare start, peak, and end times, which are determined using the X-ray flux measured by the Geostationary Operational Environmental Satellite (GOES).
The curve showing the total contribution sharply increases after the flare start time until the flare peak time, with an average increase rate of $\sim$0.4~kW~m$^{-2}$~s$^{-1}$. We also note that the rise of the total integrated losses  is delayed by a few minutes from the flare start time. The flare start time refers to the global flare, but the small patch we study evidently becomes activated sometime after the initial onset of reconnection, as flare reconnection propagates.  
The total integrated losses from our localized patch then decay at a slower rate following the impulsive peak, ($\sim$0.23~kW~m$^{-2}$~s$^{-1}$) in comparison to the sharp rise rate.
During the flaring period, the main cooling contribution comes from the \ion{Ca}{ii} and \ion{Mg}{ii} lines.
We also noticed that, at the flare peak time, the \ion{Ca}{ii}~H~\&~K and \ion{Mg}{ii}~h~\&~k lines also contribute significantly, although their contribution before the start and after the end time of the flare is less dominating compared to the \ion{Ca}{ii}~IR lines.

Moreover, during the flaring time, the radiative loss due to the Ly$\alpha$ line increases until the flare peak time, but its value stays below the \ion{Ca}{ii} and \ion{Mg}{ii} lines. At the flare footpoint, during the flare peak time, the maximum value of integrated radiative losses from Ly$\alpha$ is around $\sim$7 kW~m$^{-2}$. We note, however, that there might be a bias in this result. Our chromospheric inversion is only based on spectra from the \ion{Ca}{ii} lines. For $T\gtrsim 10$~kK, these lines rapidly lose sensitivity because of \ion{Ca}{ii} ionization into \ion{Ca}{iii} \citep{2007ApJ...670..885P}. Therefore, our models typically predict a very cold transition region in comparison with an inversion that includes lines that sample the transition region. While the effect of a colder transition region is not necessarily visible in the \ion{Ca}{ii} lines, it can have a profound effect in the Ly$\alpha$ line, which has a strong contribution from regions with temperatures ranging from 20 to 40~kK \citep{1980ApJ...242..336M, 1981ApJS...45..635V}. This could explain why the response in this line is so subdued compared to the expected stronger response \citep{2014ApJ...793...70M,2018ApJ...862...59B}. Thus, it is not only a problem of restricting the integration limits for the radiative losses to $T\le 10$~kK, but of the intrinsic structure of the reconstructed transition regions. 

\subsection{Comparison of radiative losses retrieved from observations and simulations}

In order to compare the values retrieved from observations with the simulations, we also estimated the integrated radiative losses from the atmospheric model synthesized using the radiative hydrodynamic (RADYN) code \citep{1997ApJ...481..500C, 2015ApJ...809..104A}. The RADYN simulation represents a 1D strand or field line in the flare that attempts to catch most of the microphysics that are at work in a flare. RADYN was recently merged with the code FP \citep[][]{2020ApJ...902...16A}, offering an improved treatment of the nonthermal particle propagation and dissipation. The RADYN code was employed to simulate various flaring atmospheres, which are characterized by the electron beam parameters: low-energy cutoff (E$_c$), spectral index ($\delta$), and nonthermal energy flux. The $\delta$ and E$_c$ values determine the spectral ``hardness'' of the nonthermal electron distribution. A higher $\delta$ and lower $E_{c}$ represent a softer distribution, which is a relatively larger proportion of lower energy electrons. Lower energy particles thermalize more easily in the upper chromosphere and the lower transition region compared to higher-energy electrons that can penetrate deeper. We performed a small parameter study, with a fixed $\delta = 7$, but variable E$_{c} = [10, 15, 20]$ keV. For each value of E$_c$ and $\delta$ we injected a beam with energy flux values F = [0.75$\times10^{10}$, 1.0$\times10^{10}$, 2.5$\times10^{10}$, and 5.0$\times10^{10}$]~erg~s$^{-1}$~cm$^{-2}$, for a period of $t=30$~s. Following cessation of the beam, each simulation was allowed to cool.

While generating STiC input model atmosphere, we set a threshold temperature of $40$~kK to the RADYN synthetic model atmospheres, which removed the contribution of the corona. This was required as the STiC code is not well constrained to model the coronal dynamics. 
We also investigated the influence of the temperature threshold on the radiative losses. We find that there is no significant change in the radiative loss values if the temperature threshold changed from ~40~kK up to 60~kK. This could be related to our integrated heights, which are located in the deeper chromosphere.

As an example, the temporal evolution of the temperature and velocity generated from one of the RADYN simulations is shown in Fig.~\ref{radyn_synthetic}. The obtained stratification of temperature and velocity are qualitatively similar to the inverted model parameters shown in Paper~I, but there is a steeper rise in temperature, and the velocity field has larger amplitudes in the chromosphere compared to the inverted model retrieved from observations. Although this may be a consequence of the starting atmosphere, as well as the choice of electron beam parameters, we note that node-based inversion tends to predict smooth model atmospheres as sharp gradients and very complex stratifications require a large number of nodes.

We analyzed the radiative losses in the mid-chromosphere for the different electron beam parameters and we find that the total radiative losses in that part of the atmosphere are higher for larger values of E$_c$ and energy flux. In our case, the total radiative losses obtained with E$_c=20$~keV and an energy flux of $5\times10^{10}$ erg s$^{-1}$ cm$^{-2}$ are the most  comparable to the value obtained from the inverted model atmosphere, though differences do exist. We stress that we only used a small sample of RADYN models as an illustration of how such a model-data comparison may be performed. A more comprehensive model-data comparison can follow in a future study. 

The right panel of Fig.~\ref{radyn_synthetic} displays the total integrated radiative losses obtained from different lines. The maximum value reaches around 125~kW~m$^{-2}$, which is comparable to the value ($\sim$135~kW~m$^{-2}$) obtained from the semiempirical model constructed from observations.
Compared to the inverted models, the RADYN curves have a much sharper rise and decay phase. While the RADYN model is representative of one field line, the behavior of the time evolution in Fig.~\ref{integrated_footpoint} can be understood as a superposition of events like the one seen in the rightmost panel of  Fig.~\ref{radyn_synthetic} in space and time. It is also known that the modeled flare gradual phase is too short compared to observations \citep[see][ and references therein]{2020arXiv200908407K,2022arXiv220411684A}.

Moreover, during the flare peak time in the RADYN simulations, in addition to the \ion{Ca}{ii} and \ion{Mg}{ii} lines, we also noticed a sharp rise in the net cooling rate of the L$\alpha$ line for a few seconds. In contrast to the simulation, we did not observe the sharp rise in Ly$\alpha$ losses estimated from the inferred model atmosphere, probably due to the colder and relatively poorly constrained transition region in the inversions that we mentioned earlier.

As illustrated in Fig.~\ref{radyn_synthetic}, at the beginning of the simulation ($<$~30~s), there is a sharp rise in temperature, which falls close to the temperature minimum location. The right panel in Fig.~\ref{radyn_synthetic} shows that the integrated radiative losses show no response before $t<$~30~s, even though the atmosphere is heated strongly. This behavior is due to our integrated height selection criteria (described in Section \ref{sec_rates_intro}), which result in a very small range to integrate the net radiative losses. Consequently, the integrated radiative losses do not show high values before 30~s in the lower-limited chromosphere. In this model, the radiative losses within the height (temperature) range selected actually peak \textsl{after} the cessation of flare energy injection. The rapidly cooling upper chromosphere (transition region) results in a heat flux that propagates to a greater depth. A small bubble forms, raising the temperature by up to a few thousand degrees (but below $T = 10$~kK), and elevating the electron density. This produces a strong peak in radiative losses from the lower chromosphere. Comparing the radiative losses derived from the inversion to those from the forward model, the observed losses we report may reflect heating of the lower chromosphere following cessation of the electron beam. A similar analysis with lines that are sensitive to a larger temperature range, and thus constrain the mid-upper flaring chromosphere, are essential and should be a priority in the present solar cycle.  

\section{Summary and discussion}
\label{Sec_summary}
A C2-class solar flare (SOL2019- 05-06T08:47) that occurred on May 6, 2019, was observed simultaneously in the \ion{Ca}{II}~K, \ion{Ca}{ii}~8542~\AA,\ and \ion{Fe}{i}~6173~\AA\ lines with the CRISP and CHROMIS instruments at the SST. 
We inferred an empirical flare model from the observational data by inverting all these lines simultaneously with the STiC code (see Paper~I for details). In this paper we utilized this model to estimate the net radiative losses, which ultimately are a lower-limit approximation of the chromospheric heating terms. Our analysis includes the main bound-bound chromospheric coolers: the Ly$\alpha$, \ion{Ca}{II}~H~\&~K, and \ion{Mg}{II} h~\&~k spectral lines, and the Ca IR triplet. Due to the lack of an accurate and realistically large atomic model, our calculations do not include the contribution from \ion{Fe}{ii} lines, which could be responsible for up to one half of the radiative cooling in the chromosphere \citep{1989ApJ...346.1010A}.
In order to estimate the radiative losses in the chromosphere, we integrated the contribution of all selected lines over the geometrical heights between the temperature minimum and the height below 10~kK temperature (e.g., indicated by gray area in Figure \ref{fig_loss_stratified}). Normally above 10~kK the observed chromospheric spectral lines loose sensitivity as at such high temperatures the \ion{Ca}{ii} atom is ionized into \ion{Ca}{iii}.

A novel aspect of our study is that we calculated a very high spatial-resolution map of net integrated losses around the peak of the flare and the entire time series for selected regions in the FOV of an observational dataset.
For the quiet-Sun pixels in the surroundings, the estimated net radiative losses have values around 5 kW m$^{-2}$, which are comparable to previously reported values \citep{1981ApJS...45..635V}. We find that the maximum value for a pixel located at a footpoint can go up to 175~kW m$^{-2}$.  
A similarly high value is also reported by \cite{2021A&A...647A.188D}, although in a magnetic reconnection event that took place deeper down in the photosphere and chromosphere. 
In the quiet-Sun, the main contributor to the net radiative losses are the \ion{Ca}{ii}~IR lines. Our analysis of the stratification of the net cooling rate shows that for a pixel located at a flare footpoint, the contribution from the \ion{Ca}{ii}~H~\&~K and \ion{Mg}{ii}~h~\&~k lines is also significant and comparable to the Ca~IR triplet (see Table~\ref{Table1}). The obtained temporal evolution of integrated radiative losses show that the total integrated losses rise sharply ($\sim$0.4~kW~m$^{-2}$~s$^{-1}$) before the flare peak time and decrease relatively slowly ($\sim$0.23~kW~m$^{-2}$~s$^{-1}$) after the flare peak. 

Due to sensitivity limitations of our setup in the transition region, the inverted models do not catch the sharp and hot transition region that is usually associated with simulations of flare atmospheres \citep[see e.g.,][]{2016ApJ...827...38R,2017A&A...605A.125S,2021ApJ...912..153K}. Therefore, the estimated integrated radiative losses in the Ly$\alpha$ line are very low ($\sim$7~kW~m$^{-2}$) compared to the contributions from the \ion{Ca}{ii} and \ion{Mg}{ii} lines. We find that the contribution of Ly$\alpha$ is not significant in the middle and lower chromosphere, or below a temperature of 10 kK. This low contribution of Ly$\alpha$, in the selected integrated heights, can also be related to our relatively weak flare. A significant contribution can be noticed from Ly$\alpha$ when the chromosphere is heated above 10~kK, which is possible in strong flares. A recent study of the Ly$\alpha$ contrasts by \cite{2021SoPh..296...51M} found that C-class flares, with some notable exceptions, produced a far smaller contrast compared to M- or X-class flares. It is imprtant to note, however, that those observations lacked spatial resolution, and therefore don't represent the response of a single footpoint, but the response of the global flare.

We compared our inversion results with a flaring atmosphere generated by the RADYN code. To facilitate the comparison of our semiempirical model retrieved from observations with the simulated flare models, we estimated the radiative losses using the same approach described in Section~\ref{sec_rates_intro} with the STiC code.
We find the best agreement with the RADYN simulations that have electron beam parameters E$_c=20$~keV and an energy flux of $5\times10^{10}$~erg~s$^{-1}$~cm$^{-2}$. The RADYN atmosphere has a sharper rise in temperature compared to the semiempirical atmosphere inferred from observations. Given the 1D nature of the RADYN simulation, the much wider peak in the net radiative losses derived from the observations can be understood as a superposition in space and time of events like the RADYN one. However, in both cases, the peak values of the total integrated radiative losses in the chromosphere are comparable.

In this study, the radiative losses are calculated assuming statistical equilibrium. This approximation works relatively well for \ion{Ca}{ii} and \ion{Mg}{ii} transitions \citep{2011A&A...528A...1W,2019ApJ...885..119K}. Due to large uncertainty in the parameters in the higher layers (<~log${\tau_{500}}= -5$), we did not consider the Balmer continuum in our study. However, the Balmer continuum could also contribute significantly to the radiative cooling in flares \citep{2022A&A...661A..77H}.
To further investigate the role of Ly$\alpha$ and other spectral lines that form in the upper chromosphere, we need high spatial and temporal resolution observations spanning different layers, especially in the upper chromosphere, which is now possible with ground-based (e.g., DKIST; \citealt{2015csss...18..933T} and SST) and space-based (e.g., IRIS; \citealt{2014SoPh..289.2733D}) observations. Additionally, the SPICE (\citealt{2020A&A...642A..14S}) instrument on board the Solar Orbiter mission \citep{2020A&A...642A...1M} has the capability to sample several strong emission lines formed at both the upper chromosphere, transition region, and coronal temperatures at a high spatial and temporal resolution. Future solar missions like Solar-C/EUVST \citep{2019SPIE11118E..07S} will also provide continuous spectral coverage throughout the solar atmosphere. The coordinated observations among different instruments could provide new insights to better constrain our flare models.

\begin{acknowledgements}
We would like to thank the anonymous referee for the comments and suggestions. The Swedish 1-m Solar Telescope is operated on the island of La Palma by the Institute for Solar Physics of Stockholm University in the Spanish Observatorio del Roque de los Muchachos of the Instituto de Astrof\'isica de Canarias. The Institute for Solar Physics is supported by a grant for research infrastructures of national importance from the Swedish Research Council (registration number 2017-00625).
RY is supported through the CHROMATIC project (2016.0019) funded by the Knut och Alice Wallenberg foundation.
 This project has received funding from the European Research Council (ERC) under the European Union's Horizon 2020 research and innovation program (SUNMAG, grant agreement 759548). G.S.K. acknowledges support from NASA’s Early
Career Investigator Program (Grant 80NSSC21K0460). The inversions were performed on resources provided by the Swedish National Infrastructure for Computing (SNIC) at the High Performance Computing Center at Link\"oping University. 
This research has made use of NASA’s Astrophysics Data System. We acknowledge the community effort devoted to the development of the following open-source packages that were used in this work: numpy (\url{numpy.org}), matplotlib (\url{matplotlib.org}) and sunpy (\url{sunpy.org}).
\end{acknowledgements}

\bibliographystyle{aa}
\bibliography{ref}  

\end{document}